\documentclass[twocolumn,showpacs,prl,amsmath,amssymb,floatfix]{revtex4-1}

\usepackage{amsmath,amssymb,color}
\usepackage[hidelinks]{hyperref}
\usepackage{bm}
\usepackage{epsfig}
\usepackage{psfrag}

\newcommand{\bd}{\bm}

\begin{document}

\title{
Non-perturbative renormalization group calculation of the quasi-particle
velocity and the dielectric function of graphene
}

\author{
Carsten Bauer, Andreas R\"{u}ckriegel, Anand Sharma, and
Peter Kopietz}
\affiliation{Institut f\"{u}r Theoretische Physik, Universit\"{a}t Frankfurt,  Max-von-Laue Strasse 1, 60438 Frankfurt, Germany}

\date{June 29, 2015}

 \begin{abstract}

Using a non-perturbative functional renormalization group 
 approach we calculate the  renormalized 
quasi-particle velocity $v (k)$ and the
static dielectric function $\epsilon ( k )$ of suspended graphene as functions
of an external momentum $k$.
Our numerical result for $v (k )$ can be fitted by  
$v ( k ) / v_F = A + B \ln ( \Lambda_0 / k )$,
where $v_F$ is the bare Fermi velocity, $\Lambda_0$ is an ultraviolet cutoff, and 
$A = 1.37$,  $B =0.51$
for the physically relevant value ($e^2/v_F =2.2$) of the coupling constant.
In contrast to calculations based on the static random-phase approximation, we find that
$\epsilon (k )$ approaches unity for $k \rightarrow 0$.
Our result for  $v (k )$ agrees very well with a recent measurement
by Elias {\it{et al.}} [Nat. Phys. {\bf{7}}, 701 (2011)].

\end{abstract}


\maketitle

At low energies the physical properties of graphene are dominated by
the Dirac points where the energy dispersion vanishes linearly.
In this regime many-body effects become important and can be measured
experimentally \cite{Kotov12}. 
In view of the great interest in graphene both for fundamental research and applied physics,
it is important to gain a thorough understanding of correlation effects.
Of particular interest is the renormalization of the Fermi velocity
at the Dirac points by long-range Coulomb interactions, which
has been observed experimentally in suspended graphene using cyclotron resonance \cite{Elias11}, in ARPES measurements of quasi-freestanding graphene on SiC \cite{Siegel11}, and in graphene on hexagonal boron nitride (hBN) \cite{Yu13}.
Early one-loop renormalization group (RG) calculations \cite{Gonzalez94}
predicted a logarithmic enhancement of the renormalized
Fermi velocity,
  \begin{equation}
 v_{\Lambda} / v_{F} = 1 + ( \alpha /4) \ln ( \Lambda_0 / \Lambda ) + {\cal{O}} ( \alpha^2 ),
 \label{eq:v1}
 \end{equation}
where $\Lambda$ is the infrared cutoff introduced in the RG procedure,
$\Lambda_0$ is an ultraviolet cutoff of the order of the inverse lattice spacing,
$v_{F}  =  10^6 {\rm m}/{\rm s}  $ is the bare Fermi velocity,
and $\alpha = e^2 / v_{F}$ is the relevant dimensionless coupling constant.
Because for graphene suspended in vacuum  $\alpha  \approx 2.2$
is rather large, perturbative RG calculations are not  
expected to be quantitatively accurate.

In this work, we use a functional renormalization  group 
(FRG)  approach \cite{Kopietz10,Metzner12}
to derive non-perturbative RG flow equations for the cutoff- and momentum-dependent 
velocity $v_{\Lambda} ( k ) $ and the static dielectric
function $\epsilon_{\Lambda} ( q )$ of suspended graphene.
Since we are interested in
the RG flow  of momentum-dependent {\it{functions}}, 
the field theoretical RG is not sufficient, because with this method
one can only keep track of a  
finite set of coupling {\it{constants}}.
We show here that this problem 
can be solved within the FRG formalism \cite{Kopietz10,Metzner12};
specifically, we derive two coupled integro-differential
equations for the cutoff-dependent functions
$v_{\Lambda} ( k ) $ and $\epsilon_{\Lambda} ( q )$ which are
non-perturbative in $\alpha$ and self-consistently
describe the interplay between self-energy
and screening effects.

Our starting point is the following effective Hamiltonian describing
the low-energy physics of graphene,
 \begin{eqnarray}
 {\cal{H}} & = & \sum_{p} 
 \int_{\bd{k}}
 \hat{\psi}^{\dagger}_{p} ( \bd{k} ) ( v_{p}  \bd{\sigma} \cdot \bd{k} )
 \hat{\psi}_{p} ( \bd{k} ) 
+ \frac{1}{2}  
 \int_{\bd{q}}
f_{\bd{q}} \hat{\rho}_{- \bd{q}} \hat{\rho}_{\bd{q}}, 
 \label{eq:hamiltonian}
 \end{eqnarray} 
where  $p = \pm $ labels the two Dirac points 
of the underlying tight-binding model on a honeycomb lattice, 
$v_p = p v_{F}$ is the bare Fermi velocity at Dirac point $p$,
and $\hat{\psi}_p ( \bd{k} )$ are two-component fermionic field operators
whose components are associated  with the two 
sublattices of the honeycomb lattice.
The two-component vector  $\bd{\sigma} = ( \sigma^x, \sigma^y )$
contains Pauli matrices 
acting on sublattice space, and
two-dimensional momentum integrations are denoted by
$\int_{\bd{k}} = \int \frac{ d^2 k }{ ( 2 \pi )^2}$.
The interaction in Eq.~(\ref{eq:hamiltonian})
is specified in terms of 
the Fourier transform $f_{\bd{q}} = 2 \pi e^2 / | {\bd{q}} |$
of the Coulomb interaction and the Fourier components
of the density operators, $\hat{\rho}_{\bd{q}} = \sum_p \int_{\bd{k}}
\hat{\psi}_p^{\dagger} ( \bd{k} ) \hat{\psi}_p ( \bd{k} + \bd{q} )$.
For simplicity, we
consider a given spin projection and
suppress  the spin label. We shall insert the spin-degeneracy $N_s =2$
of the electrons in Eq.~(\ref{eq:polflow}) below.

To derive FRG flow equations, we introduce a cutoff $\Lambda$  which inhibits 
the propagation of electrons with momenta $| \bd{k} | < \Lambda$.
For our purpose it is sufficient 
to work with a sharp momentum cutoff.
The regularized free propagator is then
 $ G^0_{ p , \Lambda} ( K )    =    \Theta ( k  - \Lambda  )
 \left[  i \omega - v_{p} \bd{\sigma} \cdot \bd{k}   \right]^{-1}$,
where the label $K = ( \bd{k} , i \omega )$ represents  momentum $\bd{k}$
and fermionic Matsubara frequency $i \omega$.
At some large initial cutoff $\Lambda_0$ of the order of the inverse lattice spacing the regularized 
Euclidean action of our system is
\begin{eqnarray}
 S_{\Lambda_0} [\psi , \phi ] 
& = &  - \sum_{p}\int_K \psi^{\dagger}_{p} ( K ) [G^0_{p, \Lambda_0} 
( K )]^{-1} \psi_{p}  ( K )  \nonumber
 \\
 &+  & \frac{1}{2}
  \int_{{Q}} 
 \left[  f_{\bd{q}}^{-1}  \phi ( - {Q} )   \phi ( {Q} )  
 +  2 i
  \rho ( - Q ) \phi ( Q) \right] ,
 \hspace{7mm}
 \label{eq:Sbare}
 \end{eqnarray} 
where $\psi_{p} ( K )$ is a two-component Grassmann field
and we have used a Hubbard-Stratonovich transformation
to represent the Coulomb interaction in terms of a scalar field
$\phi ( Q )$ which couples to the Fourier components
$\rho ( Q ) = \sum_p \int_K \psi_p^{\dagger} ( K ) \psi_p ( K + Q )$ of the density.
The integration symbols are $\int_K = \int_{\bd{k}}
 \int \frac{ d \omega}{2 \pi }$ and
 $\int_Q = \int_{\bd{q}}
 \int \frac{ d \bar{\omega}}{2 \pi }$. 
Here
$Q = ( \bd{q} , i \bar{\omega} )$, where $i \bar{\omega}$ is a bosonic
Matsubara frequency.
It is now straightforward to write down a formally exact 
flow equation for the generating functional $\Gamma_{\Lambda}  [\psi , \phi ]$
of the irreducible vertices describing their change
as we reduce the infrared cutoff $\Lambda_0 \rightarrow
 \Lambda$. By construction
for $\Lambda \rightarrow 0$ the flowing vertices reduce to the exact
irreducible vertices of our original
Hamiltonian (\ref{eq:hamiltonian}). As usual, we obtain an approximate
solution of this functional flow equation  
by working with a truncated form of  $\Gamma_{\Lambda}  [\psi , \phi ]$.
For our purpose, the following truncation is sufficient,
 \begin{eqnarray}
 & & \Gamma_{\Lambda} [ \psi , \phi ]   =   -  \sum_p \int_K 
\psi^{\dagger}_{p} ( K ) 
G^{-1}_{p, \Lambda}  ( K )   \psi_{p}  ( K )  
 \nonumber
 \\
 & + & 
  \frac{1}{2} \int_Q \phi ( - Q ) F^{-1}_{\Lambda} ( Q ) \phi ( Q )
 \nonumber
 \\
 &   + & \sum_{p,s} \int_K \int_Q \Gamma^{s}_{p, \Lambda} (  K , Q )
 \bar{\psi}^s_p ( K + Q ) \psi^s_p ( K )\phi ( Q ) ,
 \label{eq:truncate}
 \end{eqnarray}
where in the last term
$s = A, B$ is the sublattice label 
and $\psi^s_p ( K )$, $\bar{\psi}^s_p ( K )$ are the sublattice components of $\psi_p ( K )$ and the adjoint spinor $\psi^{\dagger}_p ( K )$.
It is convenient to express the renormalized fermionic and bosonic
propagators in terms of the corresponding self-energies as usual,
$ G^{-1}_{p, \Lambda}  ( K )    =  
 [G^{0}_{p, \Lambda}  ( K )  ]^{-1}  -  \Sigma_{p, \Lambda}  ( K ),$
and
$ F^{-1}_{\Lambda} ( Q )  =  f_{\bd{q}}^{-1} + \Pi_{\Lambda} ( Q )$.
Note that the fermionic self-energy $\Sigma_{p, \Lambda}  ( K )$   is a matrix in the sublattice labels.
Our truncation (\ref{eq:truncate}) retains only those vertices which
are already present in the bare action (\ref{eq:Sbare}). 
Although  higher order vertices with more than
three external legs are generated by the RG procedure, they are irrelevant
at the Gaussian fixed point. In fact, a simple scaling analysis 
keeping the Gaussian part of $\Gamma_{\Lambda} [ \psi , \phi ]$ invariant
shows that the renormalized vertices $\Gamma_n$ with $n$ external legs
scale as $\Lambda^{n-3}$, implying that all vertices 
with $n > 3$ external legs are irrelevant.
There are five  marginal vertices
with three external legs, corresponding to the field combinations
 $\bar{\psi}^A \psi^A \phi$,
$\bar{\psi}^B \psi^B \phi$,
 $\bar{\psi}^A \psi^B \phi$,
$\bar{\psi}^B \psi^A \phi$, and $\phi \phi \phi$.
While in our cutoff scheme
the purely bosonic $\phi \phi \phi$-vertex 
does not couple to the flow of the self-energies
$\Sigma_{p} ( K )$ and $\Pi ( Q )$,
the sublattice-changing vertices
of the type $\bar{\psi}^A \psi^B \phi$ and
$\bar{\psi}^B \psi^A \phi$
are neglected in Eq.~(\ref{eq:truncate}).
We nevertheless believe that our truncation
(\ref{eq:truncate}) is accurate because the sublattice-preserving vertices of the type
 $\bar{\psi}^A \psi^A \phi$ and
$\bar{\psi}^B \psi^B \phi$ are already finite in the bare action 
(\ref{eq:Sbare}).

\begin{figure}[t]
\includegraphics[width=80mm]{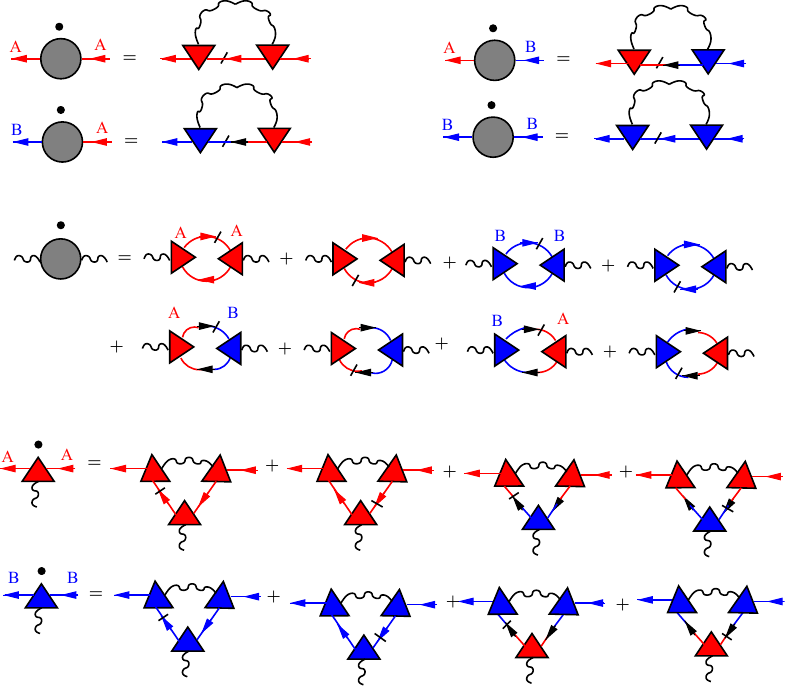}
  \caption{%
(Color online)
Diagrammatic repesentation of the FRG flow equations 
(\ref{eq:flowself}--\ref{eq:flowvertex}) for the fermionic self-energy
(first two lines), for the polarization (third and fourth line), and
the three-legged vertices (last two lines).
Solid arrows represent the exact cutoff-dependent
fermionic propagators,
wavy lines represent the corresponding bosonic propagators, and single-scale propagators
have an additional slash. 
The dot over the vertices on the left-hand side represents
the derivative with respect to the cutoff $\Lambda$.
The shaded triangles represent renormalized three-legged vertices with one bosonic and
two fermionic external legs carrying the same sublattice label. 
For clarity, we have marked some of the external legs and the vertices with
the associated sublattice labels.
}
\label{fig:Feynman}
\end{figure}

Within our  truncation and cutoff scheme, the self-energies
and three-legged vertices appearing in Eq.~(\ref{eq:truncate})
satisfy the following system of FRG flow equations,
 \begin{subequations}
 \begin{widetext}
 \begin{eqnarray}
 \partial_{\Lambda} \Sigma_{p}^{s s^{\prime}} ( K ) & = & \int_Q
  {F} ( Q) \dot{G}^{s s^{\prime}}_{ p} ( K - Q )
\Gamma^{s} _{p}  ( K-Q , Q ) 
 \Gamma^{s^{\prime}} _{p}  (  K , -Q ) ,
 \label{eq:flowself}
 \\ 
 \partial_{\Lambda} \Pi  ( Q ) & = & 
 \sum_{s s^{\prime} } \sum_{p} \int_K
 \left[ \dot{G}_{p}^{s s^{\prime}} ( K) 
 G^{s^{\prime} s }_{ p} ( K - Q )
 +   
 {G}^{s s^{\prime}}_{p} ( K) 
 \dot{G}^{s^{\prime} s }_{ p} ( K - Q )
  \right]
 \Gamma^{s} _{p}  ( K , -Q ) \Gamma^{s^{\prime}} _{p}  (  K- Q  , Q ) ,
 \label{eq:flowpol}
 \\
 \partial_{\Lambda} \Gamma^s_{p} (  K,  Q ) & = & 
 \sum_{s^{\prime}} \int_{Q^{\prime}} F ( Q^{\prime} ) \Bigl[
 \dot{G}^{s s^{\prime}}_{ p} ( K+ Q - Q^{\prime} )
 G^{ s^{\prime} s }_{ p} ( K - Q^{\prime} )
 +  {G}^{s s^{\prime}}_{ p} ( K + Q  - Q^{\prime} )
 \dot{G}^{ s^{\prime} s }_{ p} ( K - Q^{\prime} )
 \Bigr]
 \nonumber
 \\
 &   &  \times
  \Gamma^{s}_{p} (  K+ Q - Q^{\prime} , Q^{\prime} )
\Gamma^{s}_{p} (  K  , - Q^{\prime} )
 \Gamma^{s^{\prime}}_{p} (  K - Q^{\prime}  , Q ),
 \label{eq:flowvertex}
 \end{eqnarray}
\end{widetext}
 \end{subequations}
where for simplicity we have omitted the cutoff label $\Lambda$ and 
the single-scale propagator is
 \begin{eqnarray}
  \dot{G}_{ p , \Lambda} ( K )   & =  &  - \delta (k  - \Lambda  )
 \left[  i \omega - v_{p} \bd{\sigma} \cdot \bd{k}   
  - \Sigma_{p, \Lambda} ( K )     \right]^{-1}.
 \hspace{7mm}
  \end{eqnarray}
The derivation of the above FRG equations starting from the general
Wetterich  equation \cite{Wetterich93} for the 
theory defined in Eq.~(\ref{eq:Sbare})
is similar to the derivation
of the vertex expansion for mixed Bose-Fermi models discussed in 
Refs.~[\onlinecite{Schuetz05,Kopietz10}].  
A graphical representation of the flow 
equations~(\ref{eq:flowself}--\ref{eq:flowvertex})
is shown  in Fig.~\ref{fig:Feynman}.
Since we are interested in the
low-energy behavior of the Green function, 
we expand the self-energy
to linear order in the frequency,
 \begin{equation}
  \Sigma_{p, \Lambda} ( K  ) = 
  p V_{\Lambda} ( k )  
 \bd{\sigma} \cdot \bd{k}+ ( 1 - Z_{\Lambda}^{-1} ) i \omega
 + {\cal{O}} ( \omega^2 ).
 \label{eq:sigmalow2}
 \end{equation}
Here $Z_{\Lambda}$ is the wave-function renormalization factor.
Note that we do not expand the momentum dependence of the velocity correction
$V_{\Lambda} ( k)$.
For small frequencies our scale-dependent fermionic propagator is given by
 \begin{equation}
 G_{p ,\Lambda } ( K  ) 
= - \Theta ( k  - \Lambda ) Z_{\Lambda} \frac{ i \omega +  p  {v}_{ \Lambda} ( k )
 \bd{\sigma} \cdot \bd{k} }{ \omega^2 +  \xi_{\Lambda}^2 ( k ) },
 \end{equation}
where  $ \xi_{\Lambda} ( k ) = {v}_{\Lambda} ( k )  k$
is the energy dispersion at cutoff scale $\Lambda$ and
 ${v}_\Lambda ( k ) = Z_{\Lambda} [ v_F + V_{\Lambda} ( k ) ]$
is the corresponding velocity.
Moreover, we retain only the marginal part
 $\Gamma^{s}_{p,\Lambda} ( 0, 0 ) \equiv i \gamma_{\Lambda}$
of the three-point vertices.
The flow equation (\ref{eq:flowself})
for the self-energy then reduces to
 \begin{eqnarray}
 \partial_{\Lambda} \Sigma_{p , \Lambda} ( K ) & = &
 - \gamma_{\Lambda}^2 Z_{\Lambda} \int_{\bd{q} , \bar{\omega}}
 \frac{  \delta ( q - \Lambda )  f_{ \bd{k} - \bd{q}} }{  \epsilon_{\Lambda} (  \bd{k} - \bd{q} , i \omega 
 - i \bar{\omega} ) }
 \nonumber
 \\
 &  &  \hspace{10mm} \times 
\frac{ i \bar{\omega} + p {v}_{\Lambda} ( q ) \bd{\sigma} \cdot \bd{q} }{ 
 \bar{\omega}^2 + \xi_{\Lambda}^2 ( q ) } .
 \label{eq:selfflow}
 \hspace{7mm}
 \end{eqnarray}
Here we have introduced the cutoff-dependent dielectric function
 $ \epsilon_{\Lambda} ( Q ) = 1 + 
 f_{\bd{q}}
 \Pi_{\Lambda}  ( Q )$.
The FRG flow of the  polarization is given by
 \begin{eqnarray}
 & & \partial_{\Lambda} \Pi_{\Lambda}  ( Q )  = 
 - 2 N_s \gamma_{\Lambda}^2 Z_{\Lambda}^2
 \int_{\bd{k}}
\delta \left( k  - \Lambda \right) 
 \Theta \left(  | \bd{k} - \bd{q} |  - \Lambda \right)  
 \nonumber
 \\
 &   & \times \frac{ \xi_{\Lambda} (k ) +  
\xi_{\Lambda} ( | \bd{k} -  \bd{q} | ) }{  
 [ \xi_{\Lambda} ( k )  
 +    \xi_{\Lambda} ( | \bd{k} -\bd{q} | ) ]^2
 + \bar{\omega}^2 }
 \biggl[ 1 - \frac{  \bd{k}   \cdot ( \bd{k} - \bd{q} ) }{
  k  | \bd{k} - \bd{q} | }
 \biggr],
 \label{eq:polflow}
\end{eqnarray}
where we have now inserted the spin-degeneracy factor  
$N_s = 2 S +1 =2$.
It turns out that for $K=Q =0$ the flow equation (\ref{eq:flowvertex})
for the three-point vertices $\Gamma^{s}_p (K,Q) $ 
reduces 
to the Ward identity \cite{Gonzalez10} $\gamma_{\Lambda}  Z_{\Lambda} =1$, implying
a partial cancellation between self-energy and vertex corrections
in the above flow equations.
Note that this cancellation is not properly taken into account
in the random-phase approximation (RPA) where vertex corrections are assumed to be negligible \cite{Kotov08}.
Nevertheless, 
by combining the  RPA with a RG procedure
one can obtain accurate results
for the renormalized velocity of graphene \cite{Hofmann14}.

Although it is now straightforward to derive a closed system of FRG flow
equations for  $Z_{\Lambda}$ and
the two functions
 $v_\Lambda ( k )$ and
$\epsilon_{\Lambda} ( \bd{q} , i \bar{\omega} )$, 
let us neglect here the frequency dependence of the dielectric function,
$\epsilon_{\Lambda} ( \bd{q} , i \bar{\omega} ) \approx \epsilon_{\Lambda} ( q )$.
In this approximation $Z_{\Lambda} = \gamma_{\Lambda} = 1$, but the
renormalization of the Fermi velocity is non-perturbatively
taken into account. This is sufficient to obtain the correct
quantum critical scaling in gaphene \cite{Sheehy07}.
After performing one of the integrations in Eqs.~(\ref{eq:selfflow}) and
(\ref{eq:polflow}) we obtain
 \begin{subequations}
 \begin{widetext}
 \begin{eqnarray}
 \Lambda \partial_{\Lambda} v_{\Lambda} ( k ) & = &
  - \frac{e^2}{2  } \frac{\Lambda}{k} \int_0^{\pi} \frac{d \varphi}{\pi} \frac{ \cos \varphi}{
 \sqrt{ 1 - 2 ( k / \Lambda ) \cos \varphi + ( k / \Lambda )^2 }}
 \frac{1}{
 \epsilon_{\Lambda} \bigl(  \sqrt{ \Lambda^2 - 2 k \Lambda \cos \varphi + k^2 } \bigr)},
 \label{eq:flowvelocity} 
 \\
  \Lambda \partial_{\Lambda} \epsilon_{\Lambda} ( q ) & = &
  -  2 N_s e^2 \frac{q}{\Lambda} 
 \int_0^{\pi/2} \frac{d \varphi}{\pi}
  \frac{    \Theta ( 1 + \frac{q}{2 \Lambda} \cos \varphi - \frac{q}{2 \Lambda} )     }{
 \sqrt{   [ 1 + (q/2 \Lambda) \cos \varphi ]^2 - [ q/ 2 \Lambda  ]^2 }
 } \frac{   \sin^2 \varphi }{
\left[ v_{\Lambda} ( \Lambda) + 
 ( 1 + (q/ \Lambda ) \cos \varphi ) v_{\Lambda} ( \Lambda
 + q  \cos \varphi ) \right]   }.
 \hspace{7mm}
 \label{eq:floweps}
 \end{eqnarray}
 \end{widetext} 
 \end{subequations}
Note that Eq.~(\ref{eq:floweps}) has been obtained from
Eq.~(\ref{eq:polflow})  by shifting $\bd{k} \rightarrow 
 \bd{k} + \bd{q}/2$ and then introducing  elliptic coordinates.
Eqs.~(\ref{eq:flowvelocity}) and  (\ref{eq:floweps}) form
a system of coupled integro-differential equations
for the two momentum- and cutoff-dependent functions
$v_{\Lambda} ( k )$ and
$\epsilon_{\Lambda} ( q )$.
The physical renormalized velocity and the
static dielectric function 
are $v ( k ) = \lim_{\Lambda \rightarrow 0}
 v_{\Lambda} ( k )$ and
$\epsilon ( q ) = \lim_{\Lambda \rightarrow 0} \epsilon_{\Lambda} ( q )$.
We can easily recover the perturbative RG result (\ref{eq:v1})  
if we approximate
$\epsilon_{\Lambda} \approx 1$ on the right-hand side of
Eq.~(\ref{eq:flowvelocity}) and expand the integrand to leading order
in $k / \Lambda$. However, such an expansion is only valid for
$ k \ll \Lambda$, so that the physical limit $\Lambda \rightarrow 0$ at fixed
$k \neq 0$
is not accessible within this approximation.

\begin{figure}[htb]
\includegraphics[width=80mm]{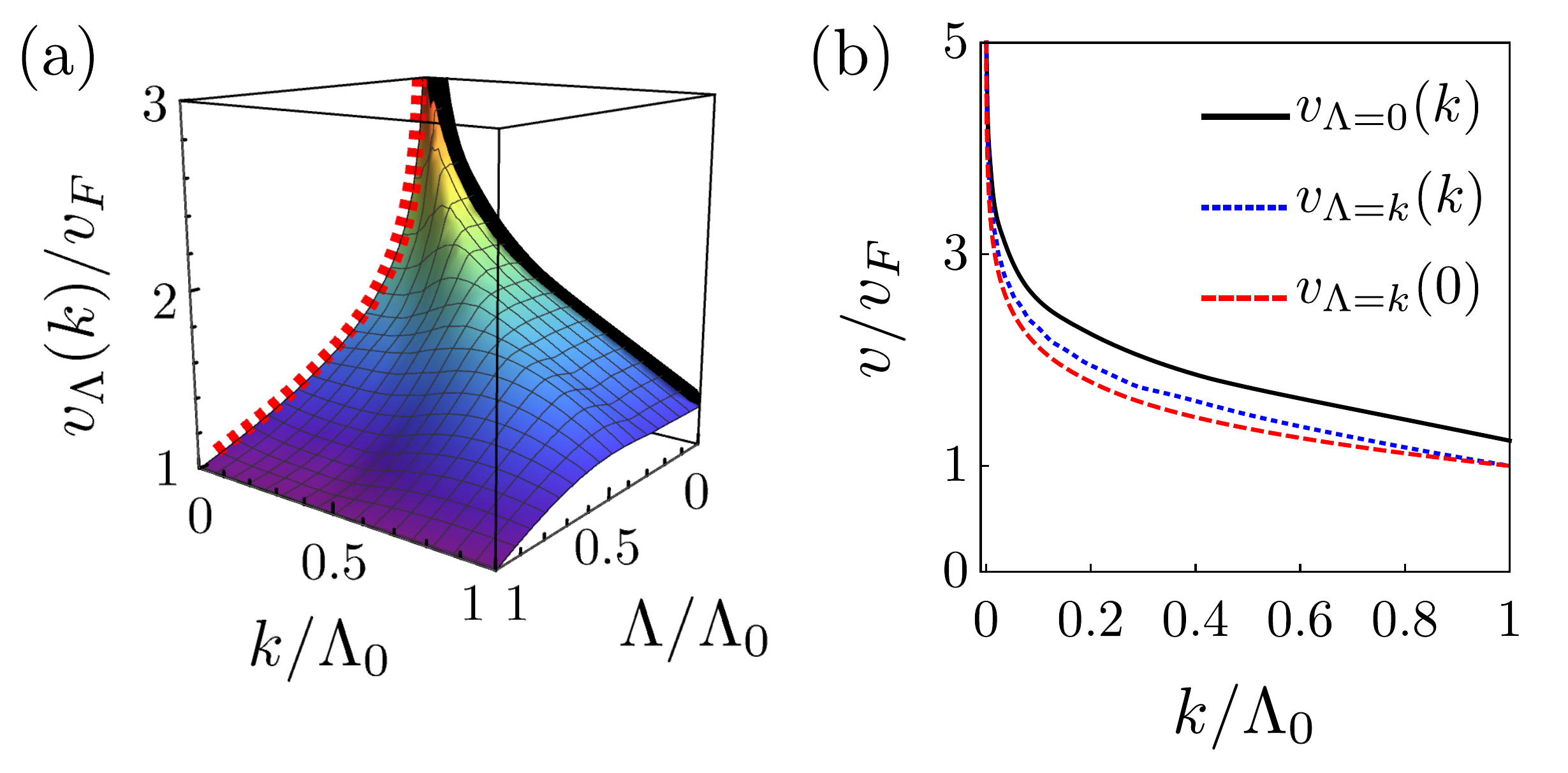}
  \caption{%
(Color online)
 (a)  Velocity
$v_{\Lambda} ( k )$ as a function of infrared cutoff $\Lambda$ and external momentum
$k$ obtained from the numerical solution of Eqs.~(\ref{eq:flowvelocity}) and (\ref{eq:floweps}).
In (b) the solid line represents  our FRG result for the physical momentum-dependent velocity 
$v ( k )  = v_{\Lambda=0} ( k )$ as a function of $k / \Lambda_0$.
For comparision, we also show  $v_{\Lambda =k} ( 0 )$ 
and  $v_{\Lambda =k } ( k )$.
\label{fig:velocitynum}
}
\end{figure}

We have solved
the FRG flow equations (\ref{eq:flowvelocity}, \ref{eq:floweps})  numerically 
without further approximations.
Note that these equations are non-perturbative in the effective coupling constant
$\alpha = e^2 /v_F$ and should be quantitatively accurate
even for large $\alpha$.
Our numerical result for $v_{\Lambda} (k) $ in
Fig.~\ref{fig:velocitynum} (a) clearly shows that the external momentum $k$
replaces  $\Lambda$ as effective infrared cutoff as soon as $\Lambda \lesssim k$.
The physical momentum-dependent velocity $v ( k ) = \lim_{\Lambda \rightarrow 0 } v_{\Lambda} ( k )$
is shown in Fig.~\ref{fig:velocitynum} (b).
For comparison, we also show the
velocities $v_{\Lambda =k} ( k )$ 
and  $  v_{\Lambda =k} (0)$; the latter can also be obtained using the field-theoretical RG
if one stops the RG flow at finite $\Lambda$ and then substitutes $\Lambda \rightarrow k$.
While
this  recipe works perfectly to leading
order in $\alpha$ where the perturbative result for $v ( k )$ 
can be obtained by replacing $\Lambda \rightarrow k$ in  Eq.~(\ref{eq:v1}),
we see
from Fig.~\ref{fig:velocitynum} (b)  that
this procedure remains approximately valid also
for the physically relevant $\alpha= 2.2$.
Our FRG result for $v ( k )$ can be fitted by
$ v ( k ) / v_F = A ( \alpha)  + B ( \alpha)  \ln ( \Lambda_0  / k )$,
where
$A (2.2) = 1.37   $ and $B (2.2) = 0.51$ for  $\alpha =2.2$.
Note that our result for $B (2.2)$ is very close to the
first order expression $\alpha /4 = 0.55$.
This is surprising, because in a perturbative expansion 
the second order correction
leads to an additional contribution
$ b_2 \alpha^2$   to  the prefactor of $\ln ( \Lambda_0 / k )$, where 
according to Mishchenko \cite{Mishchenko07} $b_2 \approx -0.14$, Vafek {\it{et al.}} \cite{Vafek08} obtained $b_2 \approx -0.3$, and Barnes {\it{et al.}} \cite{Barnes14} found $b_2 \approx -0.32$. In any case, for $\alpha = 2.2$ 
the second order correction
is substantial; our
FRG calculation suggests that in this case 
the terms of order $\alpha^2$  are to a large extent
cancelled by higher order corrections.

\begin{figure}[t]
\includegraphics[width=80mm]{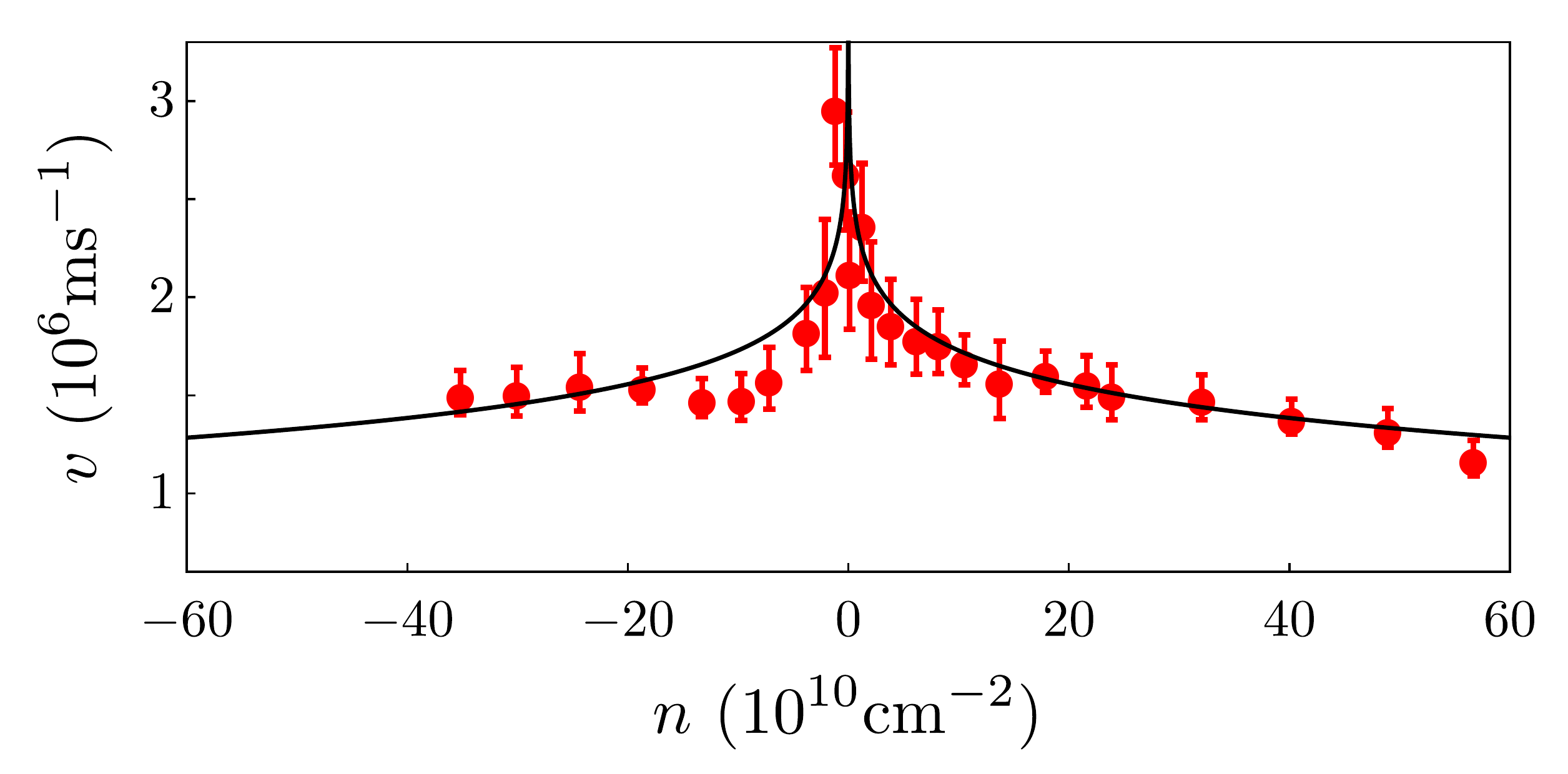}
  \caption{%
(Color online)
Comparison of our FRG result 
for the renormalized velocity $v (  k_F )$ as a function of the density
$n =  k_F^2 / \pi$ with the data from Ref.~[\onlinecite{Elias11}] (dots with error-bars).
The ultraviolet cutoff $\Lambda_0$ is fixed as described in the text.
\label{fig:velocityexp}
}
\end{figure}

To compare our results with the experimental data for the renormalized velocity
in suspended graphene \cite{Elias11},  we need to fix the value
of the ultraviolet cutoff $\Lambda_0$ which appears as  a free parameter
in our low-energy action (\ref{eq:Sbare}).
Following the usual  field-theoretical 
procedure \cite{deJuan10}, we eliminate $\Lambda_0$ in favour of some measurable observable.
Note that the authors of Ref.~[\onlinecite{Elias11}] use cyclotron resonance to measure
the quasi-particle velocity $ v ( k_F)$ at the  Fermi momentum
$k_F$  for different densities $n = k_F^2/\pi$ in slightly doped graphene. 
Assuming that the function  $v ( k )$ is approximately independent of $n$   
(which seems  to be reasonable at low densities),
we may fix $\Lambda_0$ by demanding that our result for $v ( k = k_F)$
agrees with the measured quasi-particle velocity at 
one particular value of $k_F$.
Here we choose the data point at $n \approx 49 \times 10^{10} {\rm cm}^{-2}$ 
to fix $\Lambda_0$; other choices lead to fits of similar quality.
In Fig.~\ref{fig:velocityexp} we compare the renormalized velocity obtained with
this prescription with the experimental data \cite{Elias11}.
Obviously,  in the entire range of available densities
our FRG result agrees quite well with the data. 

\begin{figure}[t]
\includegraphics[width=80mm]{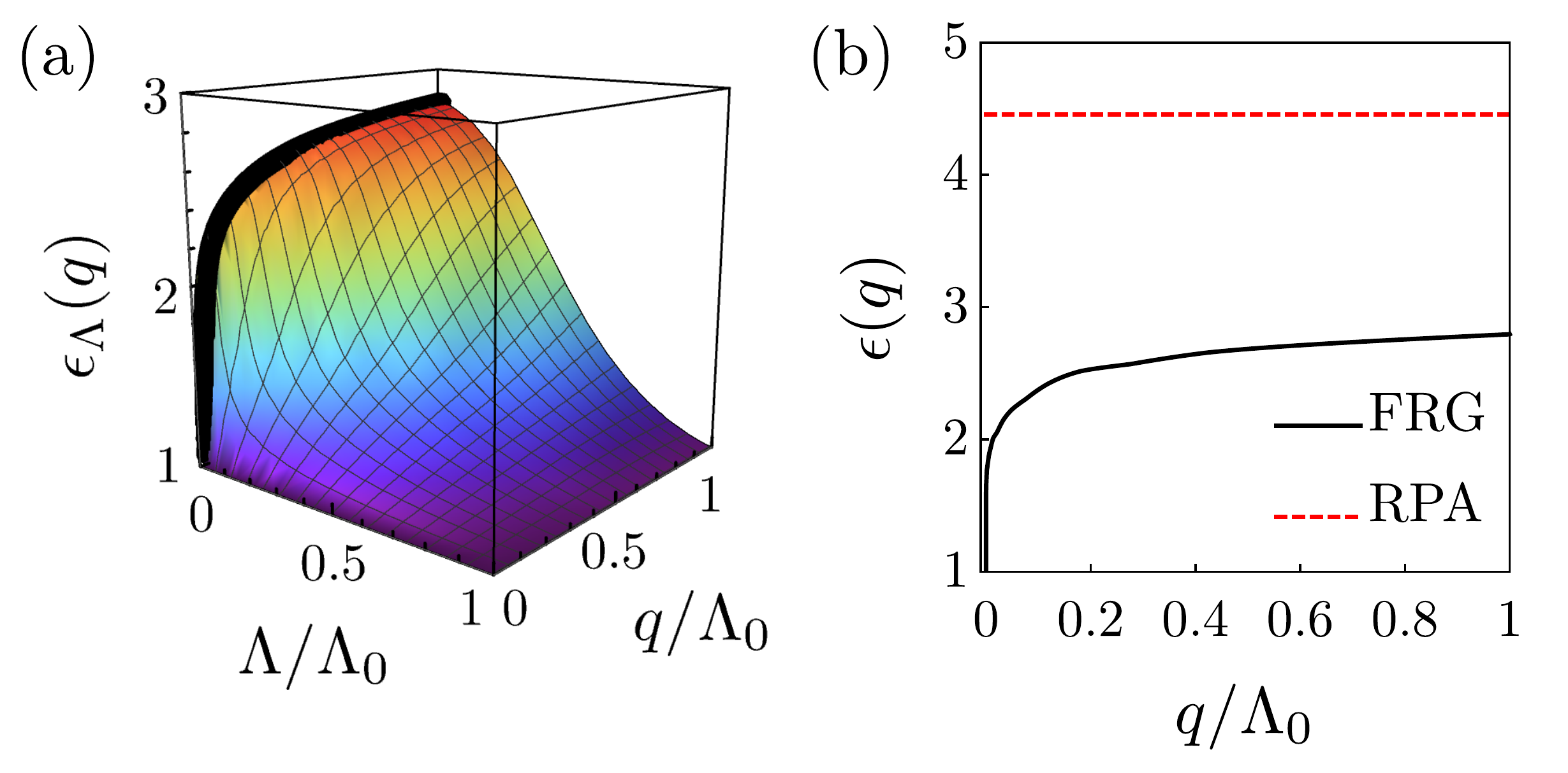}
  \caption{%
(Color online)
(a) Momentum- and cutoff-dependent 
dielectric function
$\epsilon_{\Lambda} ( q )$ obtained from the
numerical solution of the FRG flow equations (\ref{eq:flowvelocity}) and (\ref{eq:floweps}).
(b) Physical  dielectric function
 $\epsilon ( q ) = \lim_{\Lambda \rightarrow 0} \epsilon_{\Lambda} ( q )$.
The dashed line represents the RPA result for  $\Lambda_0 \rightarrow \infty$.
}
\label{fig:epsresult3d}
\end{figure}

Finally, in Fig.~\ref{fig:epsresult3d} (a) we show our numerical results for the 
momentum- and cutoff-dependent dielectric function
$\epsilon_{\Lambda} ( q )$.  Here the external momentum $q$ and the cutoff $\Lambda$
do not play the same role, because the dielectric function
is defined in terms of the bosonic self-energy $\Pi_{\Lambda} ( Q )$ while the infrared cutoff
has been  introduced in the fermionic propagator.
The physical dielectric function  $\epsilon (q ) = \epsilon_{\Lambda =0} ( q )$
is shown in Fig.~\ref{fig:epsresult3d} (b).
Note that the logarithmic divergence of the velocity $v (k)$ for small $k$
leads to the logarithmic vanishing of the static bosonic self-energy $\Pi ( q )$, so that
the corresponding dielectric function
 $\epsilon (q)$
logarithmically approaches unity
for $q \rightarrow 0$. Hence, in the static long-wavelength limit
the Coulomb interaction in suspended graphene is not screened at all. 
On the other hand, if we replace 
the flowing velocity by the bare velocity
on the right-hand side of
our flow equation (\ref{eq:floweps}) and take the limit $\Lambda_0 \rightarrow \infty$
we recover the RPA result 
$\epsilon_{\rm RPA} ( q ) = 1 + \pi N_s \alpha / 4$.

In summary, we have derived and solved non-perturbative FRG flow equations
for the  momentum- and cutoff-dependent quasi-particle velocity $v_\Lambda ( k )$ 
and the static dielectric function
$\epsilon_{\Lambda} ( q )$ of suspended graphene.
In the physical limit $\Lambda \rightarrow 0$ our result
for $v ( k )$ diverges as $  \ln ( \Lambda_0 / k )$ for $k \rightarrow 0$ and
agrees very well
with a recent experiment~\cite{Elias11}.
The dielectric function $\epsilon ( q )$ is shown to approach unity for $q \rightarrow 0$,
in contrast to the prediction of the RPA.
Our approach can be extended in several directions:
with some extra numerical effort, the frequency-dependence of the dielectric function
can be taken into account. Moreover,  we can generalize our approach to allow for
the spontaneous formation of a charge-density wave
which  transforms the system into 
an excitonic insulator \cite{Kotov12,Khveshchenko01}.

We thank A. Geim for making the data for the renormalized velocity
from Ref.~[\onlinecite{Elias11}] available to us.
Two of us (C. B. and P. K.) acknowledge the hospitality of the
physics department of the University of Florida, Gainesville, where part of this work 
has been  carried out.

\end{document}